\documentclass[prb,aps,twocolumn,draft,tightenlines,superscriptaddress]{revtex4}
\usepackage{psfig}
\usepackage{amssymb}
\begin{document}
\title{Magnon and phonon assisted tunneling in a
  high-magnetoresistance tunnel junction using Co$_{75}$Fe$_{25}$
  ferromagnetic electrodes}
\author{C. L\"u}
 \affiliation{Structure Research Laboratory, University of Science \&%
Technology of China, Academia Sinica,  Hefei, Anhui, 230026, China}
\affiliation{Department of Physics, University of Science \&%
Technology of China, Hefei, Anhui, 230026, China}
\altaffiliation{Mailing Address.}
\author{M. W. Wu}
 \thanks{Author to whom correspondence should be addressed}%
\email{mwwu@ustc.edu.cn}%
\affiliation{Structure Research Laboratory, University of Science \&%
Technology of China, Academia Sinica,  Hefei, Anhui, 230026, China}
\affiliation{Department of Physics, University of Science \&%
Technology of China, Hefei, Anhui, 230026, China}
\altaffiliation{Mailing Address.}
\author{X. F. Han}
\affiliation{State Key Laboratory of  Magnetism,  Institute of  Physics,
  Chinese Academy of Sciences,  P.O. Box 603,  Beijing 100080,  China}
\date{\today}

\begin{abstract}
Magnetoelectric properties of the spin-valve-type tunnel junction
of Ta(5\ nm)/Ni$_{79}$Fe$_{21}$ (25\ nm)/Ir$_{22}$Mn$_{78}$ (10\ nm)/Co$_{75}$Fe$_{25}$ (4\ nm)/Al
(0.8\ nm)-oxide/Co$_{75}$Fe$_{25}$ (4\ nm)/Ni$_{79}$Fe$_{21}$ (20\
nm)/Ta (5\ nm) are investigated both experimentally and theoretically.
It is shown that both magnon and phonon excitations contribute to the
tunneling process. Moreover, we show that there are two
branches of magnon with spin $S=1/2$ and 3/2 respectively. The
theoretical results are in good agreement with the experimental data.
\end{abstract}
\pacs{PACS: 75.70.Ak, 73.40.Gk, 72.10.Di, 73.50.Bk}
\maketitle


Tremendous interest has been devoted to the tunnel magnetoresistance (TMR) effect\cite{jull,miya,mood,zhang,chui,brat,daug,han1,han2,han3}
due to the high application potential in magnetic random access memory (MRAM)
and magnetic-read-head technology.\cite{sato,daug,ago,tehrani,machida}
Spin-electron transport and nanoscale magnetism in ferromagnet/insulator/ferromagnet
(FM/I/FM) junction structure play a very important role in this effect.
Up to present, although considerable progress on both experimental and
theoretical studies of TMR effect in FM/I/FM junctions has been
achieved, intrinsic magnetoelectric properties of magnetic tunnel junctions
(MTJs) as well as spin-electron transport theory have not yet been
generally reported. Therefore, a close study of these subjects are
important both for the sake of fundamental studies and for the
development of high-quality TMR devices.
\begin{figure}[htb]
  \psfig{figure=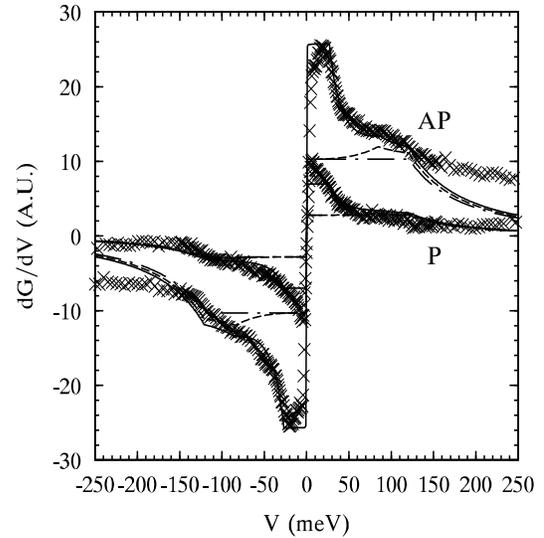,width=9.cm,height=8.cm,angle=0}
  \caption{ IET spectrum via the single-magnon-assisted
tunneling (dash-dotted curve), single-magnon and  phonon-assisted tunneling
(dashed curve) and double-magnon and single-phonon-assisted
tunneling (solid curve) are plotted as functions of voltage $V$.
The experimental data are plotted as crosses. $T=4.2$\ K.
}
\end{figure}

Recently Han {\em et al.} performed a systematic experimental
investigation of a spin-valve-type tunnel junction of
 Ta(5\ nm)/Ni$_{79}$Fe$_{21}$ (3\ nm)/Cu (20\ nm)/Ni$_{79}$Fe$_{21}$
(3\ nm)/Ir$_{22}$Mn$_{78}$ (10\ nm)/Co$_{75}$Fe$_{25}$ (4\ nm)/Al
(0.8\ nm)-oxide/Co$_{75}$Fe$_{25}$ (4\ nm)/Ni$_{79}$Fe$_{21}$ (20\
nm)/Ta (5\ nm) by measuring the tunnel current $I$, dynamic
conductance $dI/dV$ and inelastic electron tunneling (IET) spectrum
$dG/dV$ ($G=I/V$) as functions of  dc bias voltage $V$
for both parallel (P) and antiparallel (AP) alignments of the
magnetization of the two FM electrodes.\cite{han3}
By applying the magnon-assisted tunneling theory developed by
Zhang {\em et al.}\cite{zhang} they fit their $I$-$V$, TMR-$V$, and
TMR-$T$ curves with a $S=3/2$ magnon excitation $\omega_{\bf
  q}=E_m(q/q_m)^2$ for cut-off energy $E_m=121$\ meV and
$q_m=\sqrt{4\pi n}$ where $n$ is the density of atoms at an interface.
However, there is more interesting information in their IET spectrum
data which has not been fully discussed.  In particular, the contribution
of phonon-assisted tunneling to the tunneling current has not been accounted
in their model and calculations.\cite{zhang,han3}
As shown in Fig.\ 1 there are three peaks in the IET spectrum from a new sample 
obtained in this work for $V> 0$
(or $< 0$): a strong peak around 20\ meV and two small peaks around 90 and
115\ meV. The small peak around 121\ meV correspond to $E_m$ of the magnon
excitation. Another small peak around 90\ meV is identified to be
the phonon excitation in Al$_2$O$_3$ by Han {\em et al.} but has not been analyzed
theoretically. The strong peak around 20\ meV has not been clearly specified.

\begin{figure}[htb]
  \psfig{figure=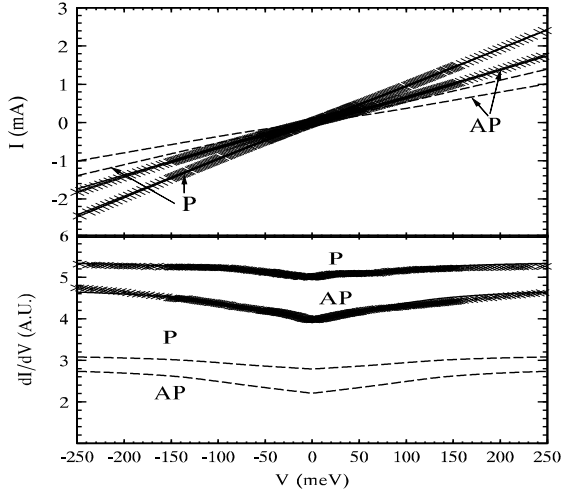,width=10.cm,height=7.cm,angle=0}
  \caption{Tunnel current $I$ and dynamic conductance
$dI/dV$ versus bias voltage.
Cross: Experimental
data; Solid curve: Theoretical results of double-magnon and
single-phonon assisted tunneling; Dashed curve:
Theoretical results of single-magnon and
single-phonon assisted tunneling. $T=4.2$\ K.
}\end{figure}

In this letter we  combine both magnon-assisted tunneling theory developed
by Zhang {\em et al.}\cite{zhang} and the phonon-assisted tunneling theory
developed by Bratkovsky\cite{brat,brat1} into one theory to analyze the
experimental data obtained in this work. This is because that the intrinsic
magnetoelectric properties of an MTJ at 
nonzero bias voltages are closely related with the inelastic scattering in the tunneling
progress for the conduction electrons. Experimental results show that the
coexisting magnon and phonon excitations are the main sources of the inelastic
scattering in an MTJ,\cite{han2} especially when the defects in the Al-O barrier
and at the interfaces between FM/I/FM layers can be neglected for a good quality
MTJ with very high-magnetoresistance and low resistance.\cite{han1}
Therefore, using a magnon- and phonon-assisted tunneling theory to evaluate the
properties of an MTJ at elevated temperature or/and at nonzero bias voltages
can achieve the desired and novel results.

In order to keep the same value of the intrinsic parameters mentioned
below for calculating self-consistently the magnetoelectric properties of an MTJ
using following formulas deduced in this work, it is necessary to use the same MTJ
for all the experimental data measurement. Therefore, a series of experimental data
was measured for a spin-valve-type MTJ of Ta (5 nm)/Ni$_{79}$Fe$_{21}$(25
nm)/Ir$_{22}$Mn$_{78}$(12 nm)/Co$_{75}$Fe$_{25}$(4 nm)/Al(0.8
nm)-oxide/Co$_{75}$Fe$_{25}$(4 nm)/Ni$_{79}$Fe$_{21}$(20 nm)/Ta(5 nm).
Such MTJs were fabricated using sputter deposition and patterned using lithographic
microfabrication technique followed by optimum heat treatment. Detailed description
was reported in previous works by Han {\em et al.}.\cite{han1}
We specify excitations for each corresponding peaks in the IET spectrum. Moreover,
we propose a second magnon excitation with $S=1/2$ and $E_m=27$\ meV which
corresponds to the large peak around 20\ meV. This excitation corresponds to the Fe
spin of the  Co$_{75}$Fe$_{25}$ electrodes.

\begin{figure}[htb]
  \psfig{figure=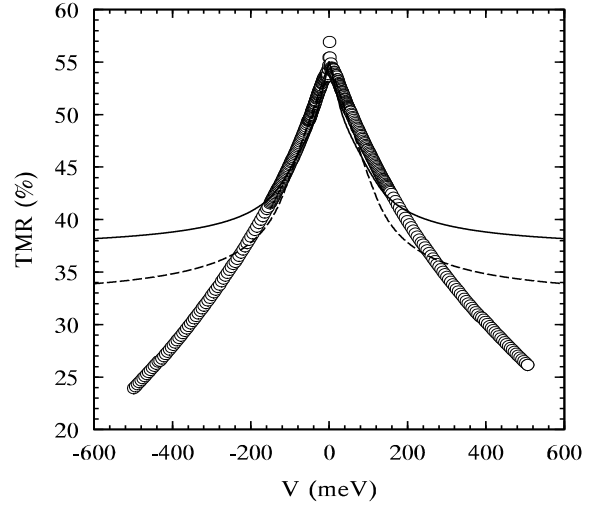,width=10.cm,height=7.5cm,angle=0}
  \caption{TMR versus bias voltage at $T=4.2$\ K. Open circles:
Experimental data; Solid curve: Calculated TMR of double-magnon and
single-phonon assisted tunneling; Dashed curve:
Calculated TMR of single-magnon and
single-phonon assisted tunneling.
}\end{figure}

We obtain for the tunneling current
$I^\gamma=I_0^\gamma+I_p^\gamma+I_{m,i}^\gamma$
where $I_0^\gamma$ is the elastic tunneling current,
$I_p^\gamma$
denotes the inelastic phonon-assisted tunneling current, and
$I_{m,i}^\gamma$
represents the inelastic magnon-assisted tunneling current
with $i=1$, 2 standing for the index of the branch of magnon.
$\gamma$ here represents P and AP. These currents are given
respectively by
$
I_0^\gamma=\frac{4\pi e^2
  V}{\hbar}[|T^d|^2+2(S_1+S_2)|T^J|^2]A^\gamma\ ,
$
with $T^d$ and $T^J$ representing direct and
spin-dependent charge transfer matrix elements between electrodes.
\begin{equation}
I_p^\gamma=\left\{
\begin{array}{ll}
\frac{e^2 CV}{\hbar}\frac{A^\gamma}{v^4}[(k_BT)^4\frac{4\pi^4}{15}+
\frac{e^4V^4}{10}],& eV<\omega_D\\
\frac{e^2 CV}{\hbar}\frac{A^\gamma}{v^4}[(k_BT)^4\frac{4\pi^4}{15}+
\omega_D^4(\frac{1}{2}-\frac{2\omega_D}{5eV})],&eV\ge\omega_D
\end{array}\right.
\label{eq2}
\end{equation}
Here $C$ is a constant from
the matrix element of electron-phonon interaction. $\omega_D$ is the
Debye frequency.
\begin{widetext}
\begin{equation}
I_{m,i}^\gamma=\left\{
\begin{array}{ll}
\frac{8\pi eS_i}{\hbar E_{m,i}}|T^J|^2 B^\gamma \left\{[\frac{1}{2}e^2V^2
-eVE_c^\gamma+\frac{1}{2}(E_c^\gamma)^2]+2eVk_BT\ln\left(\frac{1-
e^{-E_{im}/k_BT}}{1-e^{-E_c^\gamma/k_BT}}\right)\right\}
,& eV<E_{m,i}\\
\frac{8 \pi eS_i}{\hbar E_{m,i}}
|T^J|^2B^\gamma\left\{
[\frac{1}{2}(E_c^\gamma)^2+eV(E_{m,i}-E_c^\gamma)-\frac{1}{2}E_{m,i}^2)]
+2eVk_BT\ln\left(\frac{1-
e^{-E_{im}/k_BT}}{1-e^{-E_c^\gamma/k_BT}}\right)\right\}
,&eV\ge E_{m,i}
\end{array}\right.
\label{eq3}
\end{equation}
\end{widetext}
$E_c^\gamma$ was first introduced by Zhang {\em et al.}\cite{zhang} to
represent the wavelength cut-off energy of the spin wave and was
later generalized to the anisotropic one by Han {\em et al.}.\cite{han3}
In Eqs. (\ref{eq2}) and (\ref{eq3}),
$A^{\mbox{P}}=B^{\mbox{AP}}=\rho_M^2+\rho_m^2$ and
$A^{\mbox{AP}}=B^{\mbox{P}}=2\rho_M\rho_m$ with $\rho_M$ ($\rho_m$)
being the density of states in the electrodes for itinerant  majority
(minority) electrons.

\begin{figure}[hob]
  \psfig{figure=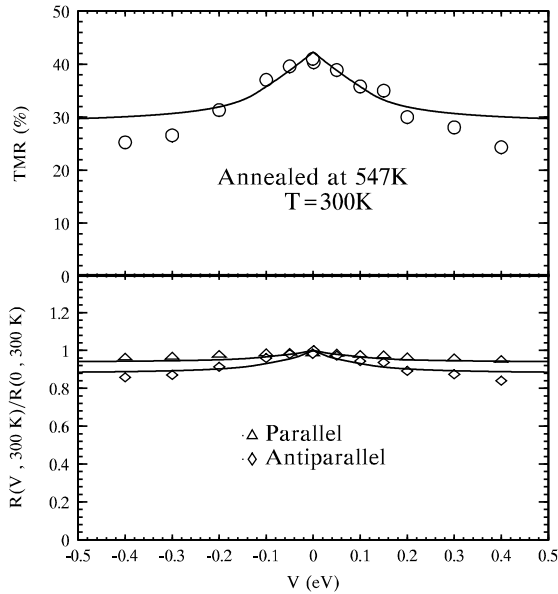,width=10.cm,height=8.5cm,angle=0}
  \caption{TMR and $R$ versus bias voltage at $T=300$\ K.
Circle: Experimental TMR data;
Triangle/Diamond: Experimental resistivity $R$ data for parallel/antiparallel
 configuration;
Solid curves: Theoretical results of single-magnon and
single-phonon assisted tunneling.
}\end{figure}

The main results of our numerical calculation are plotted in Figs.\ 1 to 6.
The parameters used in the calculation are as follows:
$R^{\mbox{AP}}(V=0)=171.5$\ $\Omega$, $R^{\mbox{P}}
(V=0)=108.1$\ $\Omega$, $\rho_M/\rho_m=2.83$ and
$|T^d|^2/|T^J|^2=13.0$. $E_c^{\mbox{AP}}=0.104$\ meV.
$E_c^{\mbox{P}}=0.009$\ meV.
The Debye frequency of Al$_2$O$_3$ is taken to be
$\omega_D=84.5$\ meV. $e^2C\rho_m^2/(\hbar v^4)=0.575$.
 In Fig.\ 1 IET spectrum via the $S=3/2$-magnon-assisted
tunneling (dash-dotted curve), $S=3/2$-magnon and  phonon-assisted tunneling
(dashed curve) and double-magnon and single-phonon-assisted
tunneling (solid curve) are plotted as functions of voltage $V$.
The experimental data are plotted as crosses in the same figure.
From the figure, one clearly sees that the small peak around
115\ meV corresponds to the magnon excitation with $S=3/2$.
If one adds phonon excitation on it, one gets the second small
peak around 90\ meV, which is close to the Debye frequency $\omega_D$.
The third strong peak is clearly related to the $S=1/2$-magnon excitation.
We point out here that as $I_{m,i}^\gamma$ is proportional to $S_i$,
therefore, the height of the magnon peaks is closely related to the
relative magnitude of spin $S_i$. For the strong peak, as $E_{m,i}$
is very small,  only $S_i=1/2$ gives the right relative  height
corresponding to the one of $S=3/2$-magnon.

\begin{figure}[htb]
  \psfig{figure=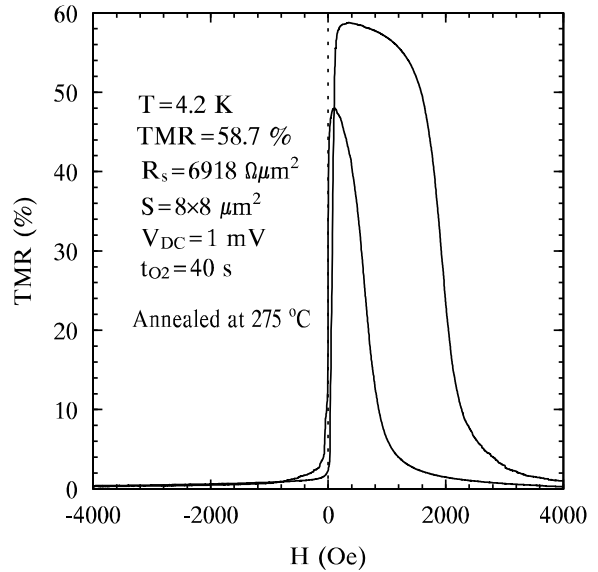,width=10.cm,height=8.5cm,angle=0}
  \caption{TMR versus external magnetic field $H$.}
  \end{figure}

In order to check if the introduction of the second magnon excitation
is still consistent with all the measurements, we plot in Figs.\ 2
and 3  the tunnel current $I$, the dynamic conductance $dI/dV$ and
TMR  as  functions of voltage $V$ at $T=4.2$\ K for
case (i) with all excitations and case (ii) with only
$S=3/2$-magnon and phonon excitations. It is interesting to seen
from the figures  that
the inclusion of the second magnon excitation
well represents the experimental results. It is noted that
for high voltage, the theoretical results deviate from the experimental
data. This is understood that for high voltages, the multi-magnon process
becomes important which nevertheless does not included in the present theory.

\begin{figure}[htb]
  \psfig{figure=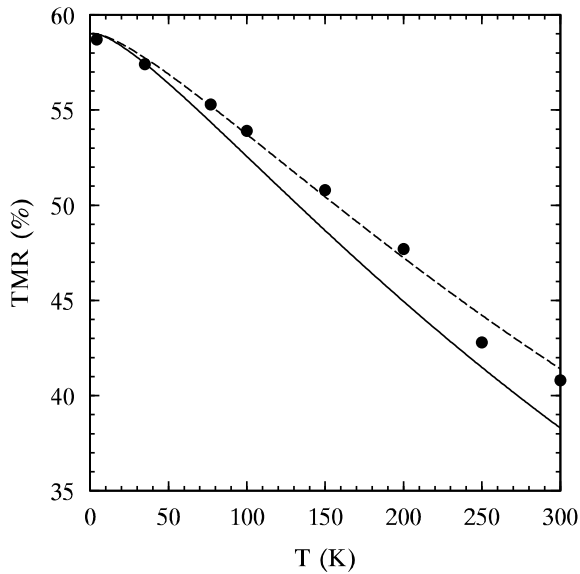,width=10.cm,height=8.5cm,angle=0}
  \caption{TMR versus temperature $T$. Dots: Experimental
data; Solid curve: Calculated TMR of double-magnon and
single-phonon assisted tunneling; Dashed curve:
Calculated TMR of single-magnon and
single-phonon assisted tunneling.}
\end{figure}

From $E_{m,i}^\gamma=3k_BT_c^i/(S_i+1)$, one has $T_c^1=156$\ K
for $S_1=1/2$ magnon. This is reasonable as the concentration of
Fe in the electrode is very small and hence the exchange
interaction between electron spins of iron is much weaker than the
bulk value. Therefore, $T_c^1$ becomes much smaller than the bulk
value. Consequently,  we predict that the strong magnon
excitation can only be seen at low temperature.

In Fig. 4 we compare our theoretical results of TMR and resistance $R$
versus bias voltage for high temperature $T=300$\ K.  As $T>T_c^1$,
one only need to consider the single magnon excitation of $S_2=3/2$
and the phonon excitation. From the figure one finds the theory
fits pretty well with the experimental data.

Finally we discuss the temperature dependence of the TMR.  As an example,
Fig.5 shows the TMR curves measured at 4.2 K for the same junction
after annealing at 548 K for one hour. The junction area is 8$\times$8 $\mu$m$^2$.
A high TMR ratio of 58.7\% was observed at 4.2 K, which was much higher than the
41\%-room temperature TMR ratio mainly due to the decrease of magnon excitations as well as the
absence of phonon excitations. In Fig.\ 6,
TMR is measured from $\sim$0\ K to 300\ K as dots. We plot our theoretical
results in the same figure for the cases of double magnon excitation (solid
curve) and single magnon excitation (dashed curve). Phonon excitation
is included in both curves. As now the temperature crosses the
Curie temperature $T_c^1$ for $S_1=1/2$-magnon excitation,
for $T>T_c^1$, only the branch of $S=3/2$-magnon participates in the
TMR. Moreover, it is understood that
 the spin-wave theory is good only for temperature far away from
the transition temperature. It is seen from the figure that the solid
curve fits the experimental data only at low temperatures. For
high temperatures, the dashed curve fits batter with the experimental
results.

In conclusion, we performed a joint experimental and theoretical investigation
of the magnetoelectric properties of the spin-valve-type tunnel junction
of Ta(5\ nm)/Ni$_{79}$Fe$_{21}$ (25\ nm)/Ir$_{22}$Mn$_{78}$ (10\ nm)/Co$_{75}$Fe$_{25}$ (4\ nm)/Al
(0.8\ nm)-oxide/Co$_{75}$Fe$_{25}$ (4\ nm)/Ni$_{79}$Fe$_{21}$ (20\ nm)/Ta (5\ nm).
We show that both magnon and phonon excitations contribute to the
tunneling process. We further point out that there are two
branches of magnon with spin $S=1/2$ and 3/2 respectively
for the MTJ using Co$_{1-x}$Fe$_{x}$ electrodes or
using Co/Al-oxide/Fe three key layers.
The theory well interpretes the experimental data.

\bigskip

MWW and XFH gratefully  acknowledge financial  support by the
``100 Person Project'' of Chinese Academy of
Sciences.  MWW is also partially supported by
Natural Science Foundation of China under Grant
No. 10247002. He would also like to thank S. T. Chui at Bartol Research Institute,
University of Delaware for hospitality. XFH  thanks support from
Natural Science Foundation of China under Grant No. 10274103
and the partial support of the State Key Project of Fundamental Research with
Grant No. 2001CB610601 of Ministry of Science and technology.
  CL would like to thank M. Q. Weng for valuable
discussion.


\begin{thebibliography}{0}
\bibitem{jull}M. Julli\'ere, Phys. Lett. {\bf 54A}, 225 (1975).
\bibitem{miya}T. Miyazaki and N. Tezuka, J. Magn. Magn. Mater. {\bf
    139}, L231 (1995).
\bibitem{mood} J. S. Moodera, L. R. Kinder, T. M. Wong, and R. Meservey,
 Phys. Rev. Lett. {\bf 74}, 3273 (1995).
\bibitem{zhang}S. Zhang, P. M. Levy, A. C. Marley, and
  S. S. P. Parkin, Phys. Rev. Lett. {\bf 79}, 3744 (1997).
\bibitem{chui}S. T. Chui, Phys. Rev. B {\bf 74}, 5600 (1997).
\bibitem{brat}A. M. Bratkovsky, Appl. Phys. Lett. {\bf 72}, 2334
  (1998).
\bibitem{daug}J. M. Daughton, J. Appl. Phys. {\bf 81}, 3758 (1997).
\bibitem{han1}X. F. Han, M. Oogane, H. Kubota, Y. Ando, and T.
Miyazaki, Appl. Phys. Lett. {\bf 77}, 283 (2000).
\bibitem{han2}X. F. Han, J. Murai, Y. Ando, H. Kubota, and
  T. Miyazaki, Appl. Phys. Lett. {\bf 78}, 2533 (2001).
\bibitem{han3} X. F. Han, A. C. C. Yu, M. Oogane, J. Murai, T. Daibou,
  and T. Miyazaki, Phys. Rev. B {\bf 63}, 224404 (2001).
\bibitem{sato}M. Sato and K. Kobayashi, IEEE Trans. Magn. {\bf 33},
3553 (1997).
\bibitem{ago}K. Tsukagoshi, B. W. Alphenaar, and H. Ago, Nature
  (London) {\bf 401}, 572 (1999).
\bibitem{tehrani} S. Tehrani, J.M. Slaughter, E. Chen, M. Durlam,
J. Shi, and M. DeHerrera, IEEE. Trans. Mags. {\bf 35}, 2814 (2000).
\bibitem{machida} K. Machida, N. Hayashi, Y. Miyamoto, T. Tamaki,
and H. Okuda, J. Magn. Magn. Mater. {\bf 235}, 201 (2001).
\bibitem{brat1}A. M. Bratkovsky, Phys. Rev. B {\bf 56}, 2344 (1997).

\end{thebibliography}
\end {document}